\pdfoutput=1

\documentclass{jors}

\begin{document}

{\bf Software paper for submission to the Journal of Open Research Software} \\

To complete this template, please replace the blue text with your own. The paper
has three main sections: (1) Overview; (2) Availability; (3) Reuse potential.

Please submit the completed paper to: editor.jors@ubiquitypress.com

\rule{\textwidth}{1pt}

\section*{(1) Overview}

\vspace{0.5cm}

\section*{Title}



FluidDyn: a Python open-source framework for research and teaching in fluid
dynamics

\section*{Paper Authors}


1. AUGIER Pierre$^a$\\
2. MOHANAN Ashwin Vishnu$^b$\\
3. BONAMY Cyrille$^a$\\

\smallskip

$^a$ Univ. Grenoble Alpes, CNRS, Grenoble INP\footnote{Institute of Engineering
Univ. Grenoble Alpes}, LEGI, 38000 Grenoble, France.\\
$^b$ Linn\'e Flow Centre, Department of Mechanics, KTH, 10044 Stockholm, Sweden.

\section*{Paper Author Roles and Affiliations}

1. Researcher, LEGI, Universit\'e Grenoble Alpes, CNRS, France\\
2. Ph.D. student, Linn\'e Flow Centre, KTH Royal Institute of Technology,
Sweden; \\
3. Research Engineer, LEGI, Universit\'e Grenoble Alpes, CNRS, France; \\

\section*{Abstract}


\href{http://fluiddyn.readthedocs.io}{FluidDyn} is a project to foster
open-science and open-source in the fluid dynamics community.  It is thought of as
a research project to channel open-source dynamics, methods and tools to do
science.
We propose a set of Python packages forming a framework to study fluid dynamics
with different methods, in particular laboratory experiments (package
\fluidpack{lab}), simulations (packages \fluidpack{fft}, \fluidpack{sim} and
\fluidpack{foam}) and data processing (package \fluidpack{image}).
In the present article, we give an overview of the specialized packages of the
project and then focus on the base package called \fluidpack{dyn}, which contains
common code used in the specialized packages.  Packages \fluidpack{fft} and
\fluidpack{sim} are described with greater detail in two companion papers,
\citet{fluidfft, fluidsim}.
With the project FluidDyn, we demonstrate that specialized scientific code can
be written with methods and good practices of the open-source community. The
Mercurial repositories are available in Bitbucket
(\url{https://bitbucket.org/fluiddyn/}). All codes are documented using Sphinx
and Read the Docs, and tested with continuous integration run on Bitbucket
Pipelines and Travis.
To improve the reuse potential, the codes are as modular as possible, leveraging
the simple object-oriented programming model of Python.
All codes are also written to be highly efficient, using C++, Cython and
Pythran to speedup the performance of critical functions.

\section*{Keywords}


Fluid dynamics research with Python. Free and open-source software, modular,
object-oriented, collaborative, efficient, tested, documented.

\section*{Introduction}


Science is mainly a collective activity.  We can go further only by
\href{https://en.wikipedia.org/wiki/%
Standing_on_the_shoulders_of_giants}{``standing on the shoulders of giants''}
(and of a huge number of technicians and scientists).  Science is a lot about
how to build new knowledge from the work of others and hence, the exchange
of ideas is a fundamental aspect.
In the last decades, we have lived through a revolution on how people exchange
ideas.
Computers of all kinds (from smartphones to HPC clusters) connected by a world
wide web are used for human communication and also for many other applications.
It has become nearly effortless to reproduce and exchange ideas and data.
The set of intangibles that grows when shared and degrades when hoarded, such as
knowledge and love, has been somehow extended.
%
The information technology revolution open doors to fantastic opportunities for
human collaboration, and of course, for science.

Web related activities alone account for at least 5\% of GDP in the USA and the
European Union\footnote{See the report by Internet Association titled
\href{https://internetassociation.org/reports/refreshing-understanding-internet-economy-ia-report/}
{``Refreshing Our Understanding of the Internet Economy''}}.
A huge amount of money (and work) is invested on the related technologies. We
are familiar with the most prominent companies involved in this dynamics
(Google, Facebook, etc.), but there are also several smaller and sometimes
lesser known organizations.  Most of these companies base part of their work on
the open-source paradigm contributing to open-source languages, libraries,
software and operating systems, while also using them --- a win-win situation
for both the corporations and the community.
This has lead to deep changes in software engineering, with a massive use of
open-source methods and tools, for example
\href{https://en.wikipedia.org/wiki/Distributed_version_control}{distributed
version control systems (DVCS)} and web-based source development platforms.

The computer performance continues to
\href{https://en.wikipedia.org/wiki/Moore\%27s_law}{increase exponentially},
now also with the help of Graphical Processing Units (GPU). This gave way to a
big boom in practical uses of data science and machine learning, which drives a
strong research on artificial intelligence. Such developments contribute to
progresses in open-source software.
To summarize, there is a strong dynamics in play around the use of computers (in
particular with the web) and this creates very efficient tools and methods
for collective work and software development.

These changes in our world also reflect in the way science is done.
Software and programming in science occupy a much bigger place than before.
%
%
The role of software in science has changed. In the past, coding was sometimes
considered as an inferior activity by some scientists.  The focus was on the
theory and the mathematical demonstration, which had to be elegant as it gets
included in the articles.  In contrast, it was normal to write crude code and to
just show the results. Nowadays, codes tend to be at the heart of research.

``Open-science'' is a new trend taking advantage of these new facts.  Pioneering
attempts are being made to do better science, improving reproducibility and
collective efficiency, by using the open-source methods and tools for science
and sharing and collaborating via the world wide web.


FluidDyn is a project to foster open-science and open-source coding in Python in
the field of fluid mechanics.
The project envisages to provide the technical framework to allow collaborative
development of tools useful for the fluid mechanics community, and to do science
with open methods.
%
We provide examples of solutions for:
\begin{itemize}
\item Good coding practice with readable and easy to comprehend Python code
(PEP~8).
\item Source control management (Mercurial) and forge (Bitbucket) simple for
the new comers.
\item Packaging and installation procedure. All packages are available from
\href{https://pypi.org/}{PyPI} and can be install simply with
\href{https://pypi.org/project/pip/}{pip}, the standard Python installation
tool.
\item Licenses: depending on the packages, we choose to use the CeCILL-B or the
CeCILL licenses. These licenses are a BSD compatible and a GPL compatible
licenses adapted to both international and French legal
matters\footnote{\url{http://www.cecill.info/licences.en.html}}.
\item Documentation produced with standard and up-to-date tools: Sphinx,
Anaconda and Jupyter. Built and hosted online at
\href{https://readthedocs.org/}{Read the Docs}. Can also be generated offline.
\item Unittest and continuous integration with Bitbucket Pipelines and Travis.
\end{itemize}
We hope that such a clean framework will facilitate contributions from
scientists in the field and that we can build together a nice, user-friendly
and efficient ecosystem specialized in research and teaching in fluid dynamics.

\section*{Implementation and architecture}


\subsection*{Organization of the code in packages}

FluidDyn was originally intended to be a single package to perform experiments
and simulations. Since a typical user may not be involved in both experiments
and simulations and also, due to increasing complexity as a virtue of rapid
development cycle made the need to decentralize FluidDyn evident.
Now, FluidDyn project hosts a number of specialized packages, namely:

\begin{itemize}
\item \fluidpack{dyn}\footnote{We use FluidDyn (with capital letters) to name the
project and \fluiddyn for the base package.}: The base package which contains
pure-python code that can be reused in scripts or in specialized FluidDyn
packages. It also contains codes for miscellaneous command-line utilities useful
for a typical fluid dynamics user.

The code of this package is presented in further detail in its documentation
(\url{https://fluiddyn.readthedocs.io}) and some prominent features are presented
in the following subsection.

\item \fluidpack{fft}~\cite[see the companion paper][]{fluidfft}: a package which
provides C++ and Python classes unifying various libraries to perform Fast Fourier
Transform (FFT) in sequential and in parallel.

\item \fluidpack{sim}~\cite[see the companion paper][]{fluidsim}: Numerically
oriented framework to run sequential and parallel Computational Fluid Dynamics
(CFD) simulations and on-the-fly post-processing for a variety of problems
(Navier-Stokes, Shallow Water, F\"oppl von K\'arm\'an equations, to name a few).
A study using \fluidpack{sim} has just been published in Physics of Fluids
\cite[]{LindborgMohanan2017}.

\item \fluidpack{lab}: Package to handle laboratory experiments. Primarily used to
communicate with various hardware devices such as motors and pumps, to handle I/O
between sensors, and to store data.
Experiments using \fluidpack{lab} have been carried out in the DAMTP fluid
laboratory \cite[Cambridge,
UK. cf.][]{LeclercqPartridgeAugierDalzielKerswell2016}, in ENS Lyon laboratory
\cite[Lyon, France,][]{salort2018} and at LEGI \cite[Grenoble,
France,][]{ISSF2016}.

\item \fluidpack{image}: Scalable image processing package which implements
various algorithms to calibrate cameras, to preprocess images, to do Particle
Image Velocimetry (PIV) and to postprocess data.
\fluidpack{image} is used to process images taken during experiments performed in the
Coriolis platform at LEGI \cite[]{ISSF2016}.

\item \fluidpack{foam}: Small package to load OpenFoam data and plot them.

\item \fluidpack{coriolis}: Small package used to carry out experiments in the
\href{http://www.legi.grenoble-inp.fr/web/spip.php?article757}{Coriolis platform}
(a large rotating platform participating in the European consortiums Euhit and
Hydralab) and open the data obtained \cite[see, for example,][]{ISSF2016}. One of
the motivations behind creating this package is to study how to use open-source to
create and share open-data.

\end{itemize}

A detailed presentation on the above packages can be found in their respective
documentations on the web and for \fluidpack{fft} and \fluidpack{sim} in the
two companion papers \cite[]{fluidfft, fluidsim}.
The code base was designed to follow Python 2.7 syntax during its genesis. Now,
it has been made forward compatible with Python 3 through the use of external
package \pack{future}.
This article will now focus on the base package \fluiddyn.

\subsection*{API of the Python library \fluiddyn}

All functions and classes defined in \fluiddyn are pure Python elements, meaning
that no extensions are implemented in \fluiddyn.  Thus the package \fluiddyn is
extremely easy to install with just a \codeinline{pip install} command and no
compilation.

The package \fluiddyn is organized into five sub-packages: 
\begin{itemize}
\item \codeinline{fluiddyn.io}: This subpackage provides utilities for
input/output to different file formats.
\item \codeinline{fluiddyn.util}: Miscellaneous utilities.
\item \codeinline{fluiddyn.calcul}: Modules for numerical computations.
\item \codeinline{fluiddyn.clusters}: Classes to launch jobs on HPC clusters.
\item \codeinline{fluiddyn.output}: Utilities to produce scientific outputs
(figures and videos).
\end{itemize}

Sub-packages \codeinline{io}, \codeinline{util} and \codeinline{calcul} are the
largest in terms of lines of code and provides the Application Programming
Interfaces (API) to support \fluidpack{sim}, \fluidpack{lab} and
\fluidpack{image}.
For the sake of brevity, we shall only describe here some of the most important
modules.

\subsubsection*{Module \codeinline{fluiddyn.util.paramcontainer}}

Sub-package \codeinline{paramcontainer} defines class
\codeinline{ParamContainer} which is a hierarchical container for any type of
parameters. As shown in
\href{http://fluiddyn.readthedocs.io/en/latest/ipynb/tuto_paramscontainer.html}{%
this tutorial}, various strengths of an object of this class include:
\begin{itemize}
\item Support containing and printing documentation on the parameters.
\item Support printing to console as XML and saving as XML, HDF5 and NetCDF
Files.
\item Evaluate data types automatically while loading from saved files.
\item Easy exploration in an interactive console.
\item Allows modification of default parameters through simple Python script
files. An error is raised if the user attempts to use an invalid parameter.
\item Graphical User Interface (GUI) frontend with PyQt.
\end{itemize}

Thus, it makes it a much more robust implementation for saving key parameters,
compared to conventional methods which rely on text or CSV files.

\subsubsection*{Module \codeinline{fluiddyn.util.serieofarrays}}

This modules provides classes to iterate over files. It is a common task in data
processing to understand how to organize a \textit{serie} of files from filenames
and formats. Typically, one has to form smaller sets of arrays contained in the
files. For example, we may have a \textit{serie} such as:

\fbox{%
\begin{minipage}{\textwidth}
  \codeinline{%
    (im1\_1.png, im1\_2.png, im1\_3.png, im2\_1.png, im2\_2.png, im2\_3.png)}
\end{minipage}
}

from which we could create different subsets like,

\fbox{%
\begin{minipage}{0.5\textwidth}
\codeinline{((im1\_1.png, im1\_2.png, im1\_3.png),\\
  (im2\_1.png, im2\_2.png, im2\_3.png))}\\
\end{minipage}
}
\fbox{%
\begin{minipage}{0.47\textwidth}
\codeinline{((im1\_1.png, im2\_1.png),\\
  (im1\_2.png, im2\_2.png),\\
  (im1\_3.png, im2\_3.png))}
\end{minipage}
}

The classes of this module allows one to do it with a quite simple and general
API as shown in
\href{http://fluiddyn.readthedocs.io/en/latest/ipynb/tuto_serieofarrays.html}{%
the tutorial on the module}.

\subsubsection*{Module \codeinline{fluiddyn.util.mpi}}

This simple module makes simultaneous sequential and MPI programming a breeze
by providing number of processes, \codeinline{nb\_proc = 1} and \codeinline{rank
= 0}, when used in sequential mode, otherwise providing the appropriate values
provided by \pack{mpi4py} package. Use of this module thwarts coding several if-else
clauses. If the program was using MPI, also defines the variable
\codeinline{comm} as an alias for the \codeinline{MPI.COMM\_WORLD} communicator.

\subsubsection*{Module \codeinline{fluiddyn.calcul.easyfft}}

Thin wrapper for an unified API using classes around the packages \pack{pyfftw}
and \pack{scipy.fftpack}.  It is very easy to perform forward and inverse Fast
Fourier Transforms (FFT) in one-, two- and three-dimensions.
The FFT can be multithreaded if the environment variable
\codeinline{OMP\_NUM\_THREADS} is defined.

\subsection*{The \fluiddyn command-line utilities}

The package \fluiddyn also provides few command-line utilities to perform
simple tasks useful for a scientist developing with Python and using the
FluidDyn packages.
\begin{itemize}

\item \codeinline{fluidinfo}

Displays important information related to software and hardware. It
includes detailed information such as currently installed FluidDyn packages,
other third-party packages, C compiler, MPI and \Numpy configuration.

\item \codeinline{fluiddump}

Utility to print the hierarchy of HDF5 and NetCDF files. It does not depend on
the NetCDF4 library.

\item \codeinline{fluidnbstripout}

Very simple layer to stripout Jupyter notebooks of output, which is useful to
keep the notebooks lightweight when included in a repository.
This tool is based on \href{https://github.com/kynan/nbstripout}{nbstripout}
but, in contrast to nbstripout,
by default the notebooks with a file name ending as `.nbconvert.ipynb' are
excluded.

\item \codeinline{fluidmat2py}

Utility to produce a strange code which is no longer Matlab and not yet Python.
This strange code is then much easier to translate into correct Python than the
original Matlab code.

\end{itemize}

\section*{Quality control}


The package \fluiddyn\ currently supplies unit tests covering around 70\% of its
code.  These unit tests are run regularly through continuous integration on Travis
CI with the most recent releases of \fluiddyn's dependencies and on Bitbucket
Pipelines inside a static \href{https://hub.docker.com/u/fluiddyn}{Docker
container}.  The tests are run using standard Python interpreter with all
supported versions.

For \fluiddyn, the code coverage results are displayed at
\href{https://codecov.io/bb/fluiddyn/fluiddyn}{Codecov}.  Using third-party
packages \pack{coverage} and \pack{tox}, it is straightforward to bootstrap the
installation with dependencies, test with multiple Python versions and combine the
code coverage report, ready for upload. It is also possible to run similar
isolated tests using \pack{tox} or coverage analysis using \pack{coverage} in a
local machine.  Up-to-date build status and coverage status are displayed on the
landing page of the Bitbucket repository.

We also try to follow a consistent code style as recomended by PEP (Python
enhancement proposals) 8 and 257. This is also inspected using lint checkers
such as \codeinline{flake8} and \codeinline{pylint} among the developers.  The
code is regularly cleaned up using the Python code formatter \codeinline{black}.

\section*{(2) Availability}
\vspace{0.5cm}
\section*{Operating system}


Windows and any POSIX based OS, such as GNU/Linux and macOS.

\section*{Programming language}


Python 2.7, 3.4 or above.



\section*{Dependencies}


We list here only the dependencies of the base package \fluidpack{dyn}.

\begin{itemize}
\item {\bf Minimum:} \Numpy, \pack{Matplotlib}, \pack{psutil}, \pack{future},
\pack{subprocess32} (for Python 2.7 only), \pack{h5py}, \pack{h5netcdf}.

\item {\bf Full functionality:} \pack{mpi4py}, \pack{Scipy}, \pack{pyfftw}
(requires FFTW library), \pack{pillow}.

\item {\bf Optional:} OpenCV with Python bindings, \pack{scikit-image}.
\end{itemize}
\section*{List of contributors}


\begin{itemize}
\item Pierre Augier (LEGI): creator of the FluidDyn project, developer of
majority of the FluidDyn packages, future-proofing with Python 3 compatibility
and documentation.

\item Ashwin Vishnu Mohanan (KTH): developer of the packages
\fluidpack{sim}, \fluidpack{fft}, \fluidpack{image} and
\fluiddyn; documentation, code coverage and continuous integration (Docker,
Bitbucket Pipelines and Travis CI).

\item Cyrille Bonamy (LEGI): developer of \fluidpack{fft}, \fluidpack{sim},
\fluidpack{image} and \fluidpack{foam}. Main maintainer of \fluidpack{foam}.

\item Antoine Campagne (LEGI): developer of \fluidpack{image} and
\fluidpack{lab}.

\item Miguel Calpe Linares (LEGI): developer of \fluidpack{sim} and
\fluidpack{lab}.

\item Julien Salort (Laboratoire de physique, ENS de Lyon): developer of
\fluidpack{lab}.

\end{itemize}

\section*{Software location:}


\begin{description}[noitemsep,topsep=0pt]
\item[Name:] PyPI
\item[Persistent identifier:] https://pypi.org/project/fluiddyn
\item[Licence:] CeCILL-B, a BSD compatible French licence. 
\item[Publisher:] Pierre Augier
\item[Version published:] 0.2.4
\item[Date published:] 02/07/2018
\end{description}

{\bf Code repository} 

\begin{description}[noitemsep,topsep=0pt]
\item[Name:] Bitbucket
\item[Persistent identifier:] https://bitbucket.org/fluiddyn/fluiddyn
\item[Licence:] CeCILL-B
\item[Date published:] 2015
\end{description}



\section*{Language}


English

\section*{(3) Reuse potential}
%
%

As a library, \fluiddyn\ has been used in the project's specialized packages.
The common code base for packages with such varied applications is a proof of
\fluiddyn's versatility and generality. It can be used in other packages outside
the FluidDyn project easily depending on the need. The command-lines tools can
be useful to all scientists working with Python. Other use cases could be:
\begin{itemize}
	\item \codeinline{ParamContainer} as a generic parameter storage
		standard.
	\item Modules within \codeinline{fluiddyn.util} for miniature tasks,
		printing in terminal with colours, getting memory usage
		information, detecting if the current Python session is inside
		IPython, creating a string with time and date, etc.
	\item Modules within \codeinline{fluiddyn.io} to read
		and save images of various formats including TIFF files with
		multiple images; easily ask user with yes/no queries; handle
		different file formats using classes, including CSV, HDF5,
		Digiflow, Dantec formats.
	\item Modules inside \codeinline{fluiddyn.clusters} subpackage to
		make job submission scriptable in HPC clusters with OAR or
		SLURM job schedulers. It could be even used for non-Python
		jobs.
\end{itemize}

There is no formal support mechanism. However, bug reports can be submitted at the
\href{https://bitbucket.org/fluiddyn/fluiddyn/issues}{Issues page on Bitbucket}.
Discussions and questions can be aired on instant messaging channels in Riot (or
equivalent with Matrix protocol)\footnote{
\url{%
  https://matrix.to/\#/\#fluiddyn-users:matrix.org}}
or via IRC protocol on Freenode at \codeinline{\#fluiddyn-users}. Discussions
can also be exchanged via the official mailing list\footnote{
\url{https://www.freelists.org/list/fluiddyn}}.

\subsection*{Conclusions}

FluidDyn is an attempt to set off collaborative dynamics based on open-source
development in fluid dynamics research.
We shall try, with this project, to explore the possibilities of open-source in
science and fluid dynamics by fully exploiting the new open-source tools and
methods.

The project is right now in a preliminary stage.  Packages are actively evolving
with interesting features and a framework for collaborative development,
packaging, documenting and testing is now well set.
However, the community around the project is currently tiny and now, we have to
work on attracting users and developers since an active community is a
criteria for success and sustenance of an open-source project.

Will people use the FluidDyn tools and collaborate through the project
FluidDyn?
There are clearly many challenges and potential barriers:
\begin{itemize}
\item Some habits in the community.
\item Knowledge and skills in the community. For example, only few people use
issue tracker and pull requests.
\item Lack of a business model for open-source software in science.  A good
quality open-source software has a cost that institutions should be willing
to fund and support.
\item Lack of recognition of the work spent in open-source, in particular for
precarious scientists.
\end{itemize}

On our side, we also have positive points.
The quality of the tools we use (Python and its scientific ecosystem,
Mercurial, Read the Docs, Jupyter, ...) is impressive.
Scientific code is done to be read and to transmit ideas. To this effect, Python
is among the best languages today.
Python starts to be a standard tool in fluid dynamics, especially used for CFD
(\href{http://dedalus-project.org/}{Dedalus},
\href{https://github.com/spectralDNS}{SpectralDNS},
\href{https://pypi.org/project/triflow/}{TriFlow},
\href{http://pylbm.readthedocs.io}{PyLBM},
\href{https://github.com/mikaem/Oasis}{Oasis}, \href{http://pyfr.org/}{PyFR},
\href{https://fenicsproject.org/}{FEniCS},
\href{http://elsa.onera.fr/Cassiopee/}{Cassiopee},
\href{https://github.com/pyCGNS}{pyCGNS}, etc.) and data analysis
(\href{http://www.openptv.net/}{OpenPTV},
\href{https://github.com/jr7/pypiv}{PyPIV}).
Moreover, we can benefit from the dynamics of Python and of emerging subjects like
deep learning and the Internet of Things.
Finally, if we manage to gather a community of users and of developers, the
collective efficiency related to open-source methods and tools can be a strong
booster.

\section*{Acknowledgements}


We thank the CNRS to finance the work of Pierre Augier while giving freedom in
terms of scientific project.
Similarly, Ashwin Vishnu Mohanan could not have been as involved in this
project without the kindness of Erik Lindborg.
We thank Antoine Campagne, Miguel Calpe Linares and Julien Salort for their
implications in the development of some of the FluidDyn packages.
We are grateful to Bitbucket for providing us with a high quality forge
compatible with Mercurial, free of cost.
Finally, we thank Gabriel Moreau and Jo\"el Sommeria for their constant will
to discuss and share their knowledge.

\section*{Funding statement}


This project has indirectly benefited from funding from the foundation Simone et
Cino Del Duca de l'Institut de France, the European Research Council (ERC)
under the European Union's Horizon 2020 research and innovation program (grant
agreement No 647018-WATU and Euhit consortium) and the Swedish Research Council
(Vetenskapsr{\aa}det): 2013-5191.
We have also been able to use supercomputers of CIMENT/GRICAD, CINES/GENCI and
the Swedish National Infrastructure for Computing (SNIC).

\section*{Competing interests}


The authors declare that they have no competing interests.




\bibliographystyle{agsm} 
\bibliography{bib}

\rule{\textwidth}{1pt}

{ \bf Copyright Notice} \\
Authors who publish with this journal agree to the following terms: \\

Authors retain copyright and grant the journal right of first publication with
the work simultaneously licensed under a
\href{http://creativecommons.org/licenses/by/3.0/}{Creative Commons Attribution
License} that allows others to share the work with an acknowledgement of the
work's authorship and initial publication in this journal.

Authors are able to enter into separate, additional contractual arrangements
for the non-exclusive distribution of the journal's published version of the
work (e.g., post it to an institutional repository or publish it in a book),
with an acknowledgement of its initial publication in this journal.

By submitting this paper you agree to the terms of this Copyright Notice, which
will apply to this submission if and when it is published by this journal.

\end{document}